\documentclass[%
reprint,
superscriptaddress,
amsmath,amssymb,
aps,
twocolumn
]{revtex4-2}

\usepackage{amsmath}
\let\oldAA\AA
\renewcommand{\AA}{\text{\normalfont\oldAA}}
\usepackage{xcolor}
\usepackage{multirow}

\usepackage{booktabs}
\usepackage{lipsum}
\usepackage{graphicx}
\usepackage{dcolumn}
\usepackage{bm}
\usepackage{amsmath}
\usepackage{amssymb}
\usepackage{dutchcal}
\usepackage{float}
\usepackage{upquote}
\begin{document}

\title{Giant photocaloric effects across a vast temperature range in ferroelectric perovskites}

\author{Riccardo Rurali}
    \affiliation{Institut de Ci\`encia de Materials de Barcelona, ICMAB--CSIC, Campus UAB, 08193 Bellaterra, Spain}

\author{Carlos Escorihuela-Sayalero}
    \affiliation{Group of Characterization of Materials, Departament de F\'{i}sica, Universitat Polit\`{e}cnica de Catalunya, 
    Campus Diagonal-Bes\`{o}s, Av. Eduard Maristany 10--14, 08019 Barcelona, Spain}
    \affiliation{Research Center in Multiscale Science and Engineering, Universitat Politècnica de Catalunya, 
    Campus Diagonal-Bes\`{o}s, Av. Eduard Maristany 10--14, 08019 Barcelona, Spain}

\author{Josep Llu\'is Tamarit}
    \affiliation{Group of Characterization of Materials, Departament de F\'{i}sica, Universitat Polit\`{e}cnica de Catalunya, 
    Campus Diagonal-Bes\`{o}s, Av. Eduard Maristany 10--14, 08019 Barcelona, Spain}
    \affiliation{Research Center in Multiscale Science and Engineering, Universitat Politècnica de Catalunya, 
    Campus Diagonal-Bes\`{o}s, Av. Eduard Maristany 10--14, 08019 Barcelona, Spain}

\author{Jorge \'I\~niguez-Gonz\'alez}
    \affiliation{Luxembourg Institute of Science and Technology (LIST), Avenue des Hauts-Fourneaux 5, L-4362 Esch/Alzette, 
    Luxembourg}
    \affiliation{Department of Physics and Materials Science, University of Luxembourg, 41 Rue du Brill,
             L-4422 Belvaux, Luxembourg}

\author{Claudio Cazorla}
    \affiliation{Group of Characterization of Materials, Departament de F\'{i}sica, Universitat Polit\`{e}cnica de Catalunya, 
    Campus Diagonal-Bes\`{o}s, Av. Eduard Maristany 10--14, 08019 Barcelona, Spain}
    \affiliation{Research Center in Multiscale Science and Engineering, Universitat Politècnica de Catalunya, 
    Campus Diagonal-Bes\`{o}s, Av. Eduard Maristany 10--14, 08019 Barcelona, Spain}

\begin{abstract}
	Solid-state cooling presents an energy-efficient and environmentally friendly alternative to traditional refrigeration 
	technologies that rely on thermodynamic cycles involving greenhouse gases. However, conventional caloric effects face 
	several challenges that impede their practical application in refrigeration devices. Firstly, operational temperature 
	conditions must align closely with zero-field phase transition points; otherwise, the required driving fields become 
	excessively large. But phase transitions occur infrequently near room temperature. Additionally, caloric effects 
	typically exhibit strong temperature dependence and are sizeable only within relatively narrow temperature ranges. 
	In this study, we employ first-principles simulation methods to demonstrate that light-driven phase transitions in 
	polar oxide perovskites have the potential to overcome such limitations. Specifically, for the prototypical ferroelectric 
	KNbO$_{3}$ we illustrate the existence of giant \textit{photocaloric} effects induced by light absorption ($\Delta S_{\rm PC} 
	\sim 100$~J~K$^{-1}$~kg$^{-1}$ and $\Delta T_{\rm PC} \sim 10$~K) across a vast temperature range of several hundred 
	Kelvin, encompassing room temperature. These findings are expected to be generalizable to other materials exhibiting 
	similar polar behavior. 
\end{abstract}

\maketitle

\begin{figure}[t]
\centerline{
\includegraphics[width=1.00\linewidth]{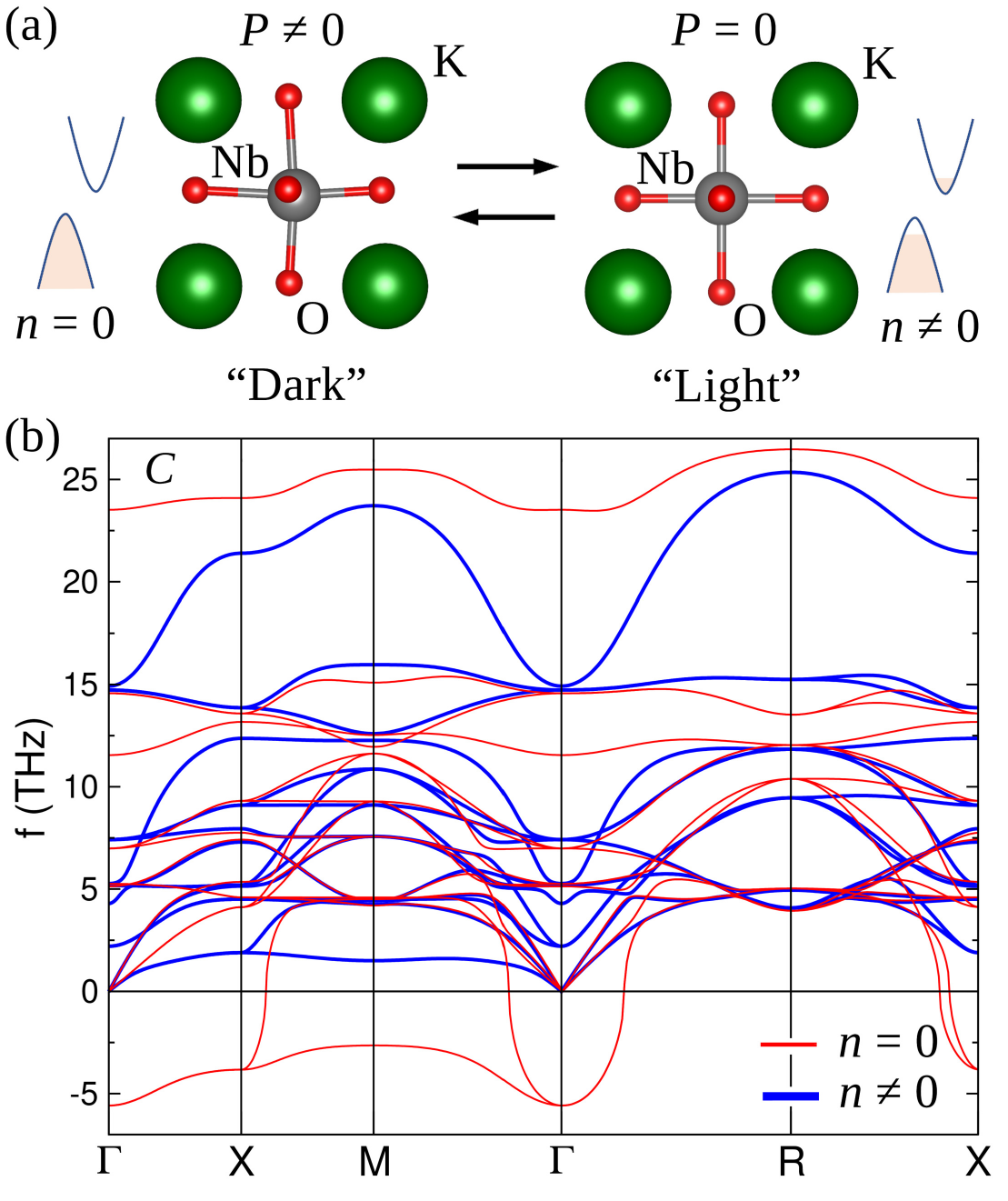}}
\caption{\textbf{Light-driven phase transition mechanism in ferroelectric KNbO$_{3}$.}
        (a)~Upon photoexcitation, electrons are promoted to the conduction band and ferroelectric KNO transforms into a 
	nonpolar cubic phase, $C$. (b)~Vibrational phonon spectrum of KNbO$_{3}$ in the nonpolar $C$ phase at $T = 0$~K 
	conditions in the absence (red lines) and presence (blue lines) of photoexcited electrons ($\overline{n} = 2.69 
	\cdot 10^{21}$~cm$^{-3}$).}
\label{fig1}
\end{figure}

Solid-state cooling is an environmentally friendly and highly energy-efficient technology that harnesses caloric effects 
in materials for refrigeration purposes \cite{cool1,giant3}. Through the application of external fields to caloric 
materials, large reversible entropy and temperature changes ($\Delta S \sim 10$--$100$~J~K$^{-1}$~kg$^{-1}$ and $\Delta T 
\sim 1$--$10$~K) can be achieved, seamlessly integrated into cooling cycles without reliance on greenhouse gases. Particularly, 
polar materials undergoing ferroelectric to paraelectric phase transitions under small electric fields are well-suited for 
solid-state cooling applications based on the electrocaloric (EC) effect \cite{ec1,ec2,ec3,ec4,ec5}, which can be miniaturized 
and coherently integrated into electronic circuits.

For caloric phenomena to be integrated in practical applications, the involved zero-field phase transition points must be 
close to room temperature. Otherwise, the necessary driving fields may reach impractically high levels, creating an energetically 
unfavorable scenario. Furthermore, the materials performance may suffer due to factors such as leakage/eddy currents, 
dielectric/magnetic losses, and mechanical fatigue. Unfortunately, only a limited number of compounds exhibit phase transitions 
near room temperature, which limits the range of solid-state refrigeration applications. Additionally, caloric effects 
typically exhibit strong temperature dependence and are substantial only within relatively narrow temperature ranges.

\begin{figure*}[t]
\centerline{
\includegraphics[width=0.90\linewidth]{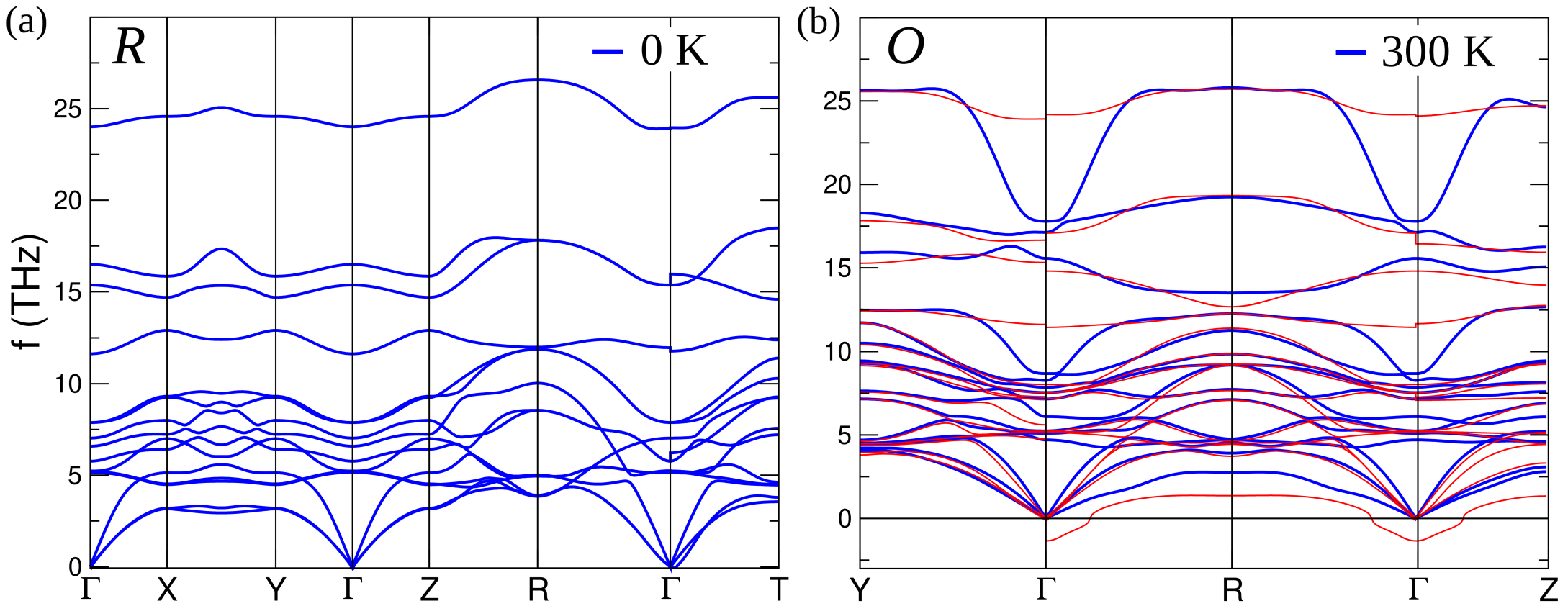}}
\caption{\textbf{Phonon spectrum of KNbO$_{3}$ in the ferroelectric $R$ and $O$ phases at ``dark'' conditions.}
         The $R$ phase is dynamically stable at zero temperature (a) whereas the $O$ phase is not as shown 
	 by the thin red lines in (b). The $T$-renormalized phonon spectrum of the $O$ phase renders dynamical 
	 stability around room temperature (blue lines) in agreement with the experimental observations 
	 \cite{wang20,shirane54}.}
\label{fig2}
\end{figure*}

In this study, we propose a solution to the described caloric materials challenges by focusing on polar to nonpolar phase 
transitions triggered by light absorption \cite{paillard19}. Specifically, we theoretically demonstrate the existence of giant 
\textit{photocaloric} (PC) effects ($\Delta S_{\rm PC} \sim 100$~J~K$^{-1}$~kg$^{-1}$ and $\Delta T_{\rm PC} \sim 10$~K) in the 
archetypal ferroelectric KNbO$_{3}$ (KNO). These PC effects persist across a vast temperature range spanning several hundred 
Kelvin, including room temperature, and are comparable to state-of-the-art EC effects \cite{ec1,ec2,ec3,ec4,ec5}. The substantial 
anharmonicity of the paraelectric phase, naturally occurring at high temperatures ($ \gtrsim 700$~K \cite{wang20,shirane54}), 
coupled with its stabilization through light irradiation, are the primary factors responsible for the unravelled giant PC effects.

We performed density functional theory (DFT) calculations \cite{vasp} with the projector augmented wave method \cite{paw} 
and generalized-gradient PBEsol approximation \cite{pbesol} (energy cutoff of $650$~eV). The electrons treated as valence 
were: K 4$s^{1}$ 3$p^{6}$ 3$s^{2}$, Nb 4$d^{4}$ 5$s^{1}$ 4$p^{6}$, and O 2$p^{4}$ 2$s^{2}$. The first Brillouin zone (BZ) 
was sampled with a $12 \times 12 \times 12$ \textbf{k}-points grid and the atomic positions were optimized until the atomic 
forces were smaller than $0.5$~meV~\AA$^{-1}$. Electric polarizations were calculated within a linear approximation using 
Born effective charges \cite{born}. Photoexcitation was mimicked by constraining the partial occupancies of the electronic 
orbitals through adjustment of the Fermi distribution smearing. This effective DFT approach is effectively equivalent to 
those employed in previous works where the concentration of electron-hole pairs was constrained via the introduction of two 
adjustable quasi-Fermi levels \cite{paillard19,monserrat20}. The second-order interatomic force constants were computed by 
finite differences with Phonopy \cite{phonopy} using $5 \times 5 \times 5$ supercells for the polar phases and $4 \times 4 
\times 4$ supercells for the nonpolar phase. Thermal expansion effects were appropriately accounted for with the quasi-harmonic 
approximation (QHA) \cite{qha}. The DynaPhoPy code \cite{dynaphopy} was used to calculate the anharmonic lattice dynamics of 
the room-temperature polar phase ($T$-renormalized phonons) from \textit{ab initio} molecular dynamics (AIMD) simulations. 
DFT Gibbs free energies, $G$, and entropies, $S$, were computed with the QHA method considering $T$-renormalized 
phonons for the phases that are dynamically unstable at zero-temperature conditions (``DFT-rQHA''). AIMD simulations were 
carried out in the $(N, V, T)$ ensemble at $T = 300$~K using Nosé-Hoover thermostats. A large simulation cell containing 
$N = 625$ atoms was employed with periodic boundary conditions applied along the three Cartesian directions. Newton's equations 
of motion were integrated using the customary Verlet's algorithm with a time step of $1.5 \cdot 10^{-3}$~ps. $\Gamma$-point 
sampling was used for BZ integration and the total duration of the AIMD simulations was of $60$~ps.  

The ground state of KNO is a polar rhombohedral phase ($R3m$, $R$). At temperatures of $220 \le T \le 470$~K, KNO stabilizes 
in a different polar orthorhombic phase ($Amm2$, $O$) \cite{wang20,shirane54}. Upon absorption of above-bandgap light, some 
electrons in KNO are photoexcited and promoted from the valence to the conduction band, where they gain increased mobility 
and become delocalized throughout the crystal. These carriers effectively screen the long-range dipole-dipole interactions 
that are key to the existence of polar order in the ferroelectric phase. Consequently, the nonpolar cubic phase ($Pm\overline{3}m$, 
$C$), which is stable at temperatures $T \gtrsim 700$~K under ``dark'' conditions \cite{wang20,shirane54}, becomes the ground 
state under ``light'' conditions (Fig.~\ref{fig1}a). 

Our zero-temperature DFT calculations reveal that the nonpolar $C$ phase becomes vibrationally stable at photoexcited electronic 
densities exceeding $\overline{n} = 2.69 \cdot 10^{21}$~cm$^{-3}$ (Fig.~\ref{fig1}b). Importantly, our DFT simulations 
demonstrate that both the polar $R$ and $O$ phases spontaneously relax into the nonpolar $C$ phase under sufficiently high light 
irradiation (or equivalently, density of photoexcited electron-hole pairs). Thus, under such photoexciting conditions the polar 
phases $R$ and $O$ are unstable rather than metastable. A similar light-induced phase transition has been shown to produce 
sizeable lattice thermal conductivity changes, with possible applications in energy scavenging and phonon-based logic 
\cite{photophononic}.

In this study, above-bandgap light absorption is crucial since it drives the promotion of charge carriers from the 
valence to the conduction band. Hence, in practice the light frequency range should be in the ultraviolet region 
(i.e., $h\nu \ge E_{g}^{\rm KNO} \approx 3$~eV \cite{saifu67}). An essential parameter for the practical relevance 
of the described photoinduced phase-transition mechanism is the light penetration length, denoted as $\Delta z$. 
If $\Delta z$ is too small, it could limit technological applications since the light-induced phase transition may not 
occur homogeneously unless $\Delta z \sim L$, where $L$ represents a characteristic size of the sample. Based on 
experimental optical absorption data for KNO \cite{saifu67}, we estimate $\Delta z \approx 30$~$\mu$m, which significantly 
exceeds the usual thickness of oxide perovskite thin films ($\sim 0.1$--$1.0$~$\mu$m). Therefore, for practical 
implementations of the \textit{photocaloric} phenomena revealed in this study, films are the most advisable 
systems.

\begin{figure}[t]
\centerline{
\includegraphics[width=1.00\linewidth]{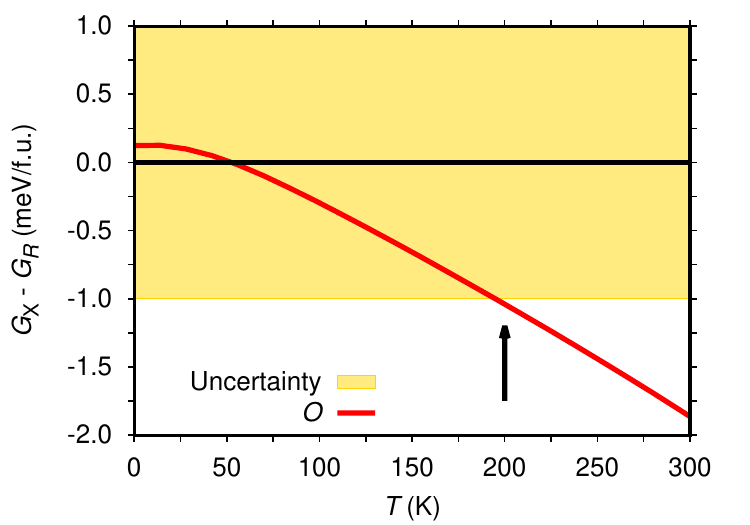}}
\caption{\textbf{DFT-rQHA prediction of the $T$-induced $R \to O$ phase transition in KNbO$_{3}$.} The Gibbs 
	free energy of the $R$ and $O$ phases cannot be distinguished in the temperature range $0 \leq T \lesssim 200$~K, 
	taking into account the numerical uncertainties of $1$~meV per formula unit. The polar $O$ phase is predicted to 
	become stable at temperatures $\gtrsim 200$~K (arrow) in fair agreement with the experiments \cite{wang20,shirane54}.}
\label{fig3}
\end{figure}

According to our first-principles DFT calculations, the polar $R$ phase is characterized by a lattice parameter of $4.034$~\AA, 
rhombohedral angle of $89.89^{\circ}$ and ferroelectric polarization of $P = 35.05$~$\mu$C~cm$^{-2}$. These results compare 
remarkably well with the corresponding experimental values ($T = 230$~K) of $4.016$~\AA, $89.83^{\circ}$ and 
$30$--$40$~$\mu$C~cm$^{-2}$, respectively \cite{wang20,kno1,kno2}. For the polar $O$ phase, we estimated the unit cell parameters 
$a = 3.996$~\AA, $b = c = 4.053$~\AA, $\alpha = 90.18^{\circ}$, and $\beta = \gamma = 90^{\circ}$, and a ferroelectric 
polarization of $P = 38.53$~$\mu$C~cm$^{-2}$ along the pseudocubic direction $[011]$. These calculated values are also in very 
good agreement with the corresponding experimental results ($T = 295$~K) $3.973$~\AA, $4.035$~\AA, $90.27^{\circ}$, and $\approx 
40$~$\mu$C~cm$^{-2}$ \cite{wang20,kno1,kno2}. Therefore, the employed DFT approach offers a robust description of KNO.

Figure~\ref{fig2} displays the vibrational phonon spectrum of the polar $R$ and $O$ phases at ``dark'' conditions. In the 
case of the $O$ phase, the phonon frequencies were renormalized to account for thermal effects at room temperature using 
a normal-mode decomposition method \cite{dynaphopy}. Our computational results accurately replicate the dynamical stability 
observed in experiments for the polar $R$ and $O$ phases at low and room temperature, respectively \cite{wang20,shirane54}. 

\begin{figure*}[t]
\centerline{
\includegraphics[width=0.90\linewidth]{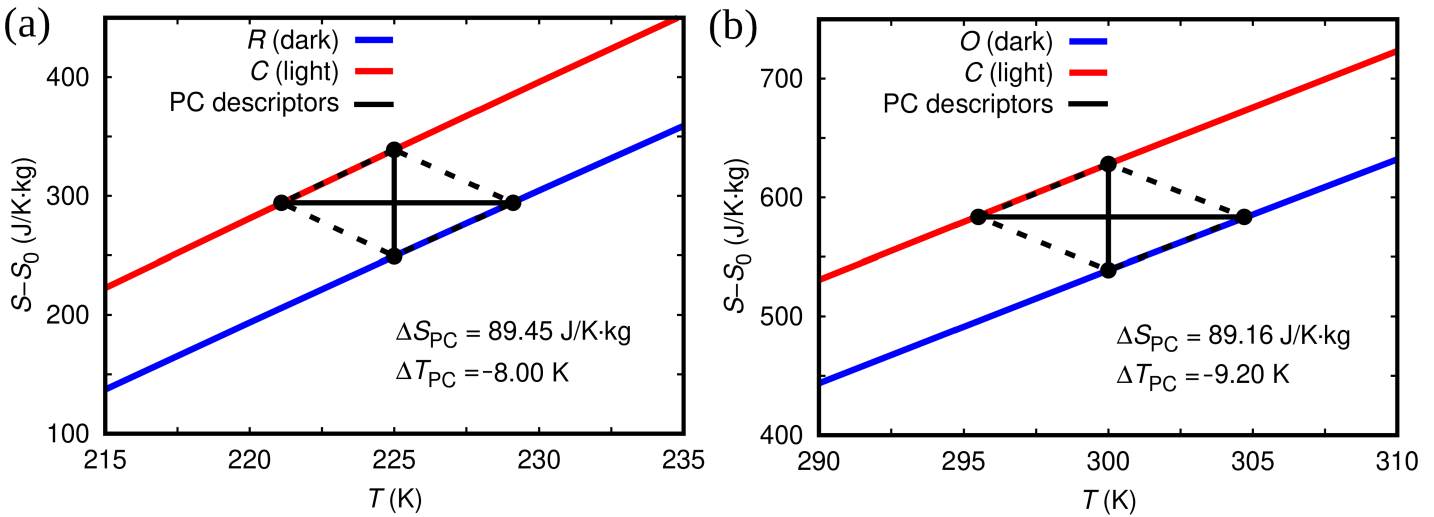}}
\caption{\textbf{Evaluation of photocaloric effects in KNbO$_{3}$.} 
	(a)~At low temperatures, for the polar $R$ (dark) and nonpolar $C$ (light) phases. (b)~Around room-temperature, 
	for the polar $O$ (dark) and nonpolar $C$ (light) phases. $\Delta S_{\rm PC}$ and $\Delta T_{\rm PC}$ can be 
	straightforwardly determined from the shown entropy curves (black solid lines) as done in quasi-direct calorimetry 
	experiments \cite{quasi1,quasi2,quasi3}. ``Light'' conditions correspond to a photoexcited electronic density of 
	$\overline{n} = 2.69 \cdot 10^{21}$~cm$^{-3}$.}
\label{fig4}
\end{figure*}

The Gibbs free energy ($G$) of the polar $R$ and $O$ phases was estimated with a numerical uncertainty of approximately 
$1$~meV per formula unit (f.u.) as a function of temperature (i.e., based on the size of the employed supercells and other
technical parameters), using the DFT-rQHA approach described earlier. These $G$ functions enable the prediction of the 
temperature at which the phase transition $R \to O$ occurs, $T_{t}$, which is determined from the condition $G_{O} (T_{t}) 
= G_{R} (T_{t})$. Figure~\ref{fig3} presents our DFT-rQHA Gibbs free energy results. We found that, considering the numerical 
uncertainties of $1$~meV/f.u. and given the proximity in energy (near-degeneracy) of the two polymorphs, the Gibbs free 
energy of the $R$ and $O$ phases could not be distinguished in the temperature range $0 \leq T \lesssim 200$~K. However, 
at higher temperatures it is clearly appreciated that $G_{O} < G_{R}$. Consequently, from our calculations it can be assuredly 
concluded that the polar $O$ phase becomes stable at temperatures above $\approx 200$~K (Fig.~\ref{fig3}), which is in 
reasonable agreement with the experimental transition temperature of $220$~K \cite{wang20,shirane54}.

Subsequently, we computed the entropy, $S = -\partial G / \partial T$, of the $R$, $O$, and $C$ phases as a function of 
temperature and light irradiation conditions, denoted as $S(T, x)$ (where $x = d, l$ stand for ``dark'' and ``light'' conditions,
respectively). The resulting $S(T, x)$ curves exhibit smooth behavior, as explicitly shown in Fig.~\ref{fig4}. Analogous 
to quasi-direct calorimetry experiments \cite{quasi1,quasi2,quasi3}, the determination of the \textit{photocaloric} isothermal 
entropy, $\Delta S_{\rm PC}$, and adiabatic temperature changes, $\Delta T_{\rm PC}$, can be readily inferred from these 
entropy curves \cite{escorihuela24}. In particular, we have that $\Delta S_{\rm PC} = S(T, l) - S(T, d)$ and $\Delta T_{\rm PC} 
= T_{0}(S, l) - T(S, d)$, where $T_{0}$ fulfills the condition $S(T_{0}, l) = S(T, d)$ (black solid lines in Fig.~\ref{fig4}). 
We recall that under sufficiently high light irradiation (or equivalently, for photoexcited electronic densities $\ge 
\overline{n} = 2.69 \cdot 10^{21}$~cm$^{-3}$) the polar $R$ and $O$ phases spontaneously relax into the nonpolar $C$ phase. 

\begin{figure*}[t]
\centerline{
\includegraphics[width=0.90\linewidth]{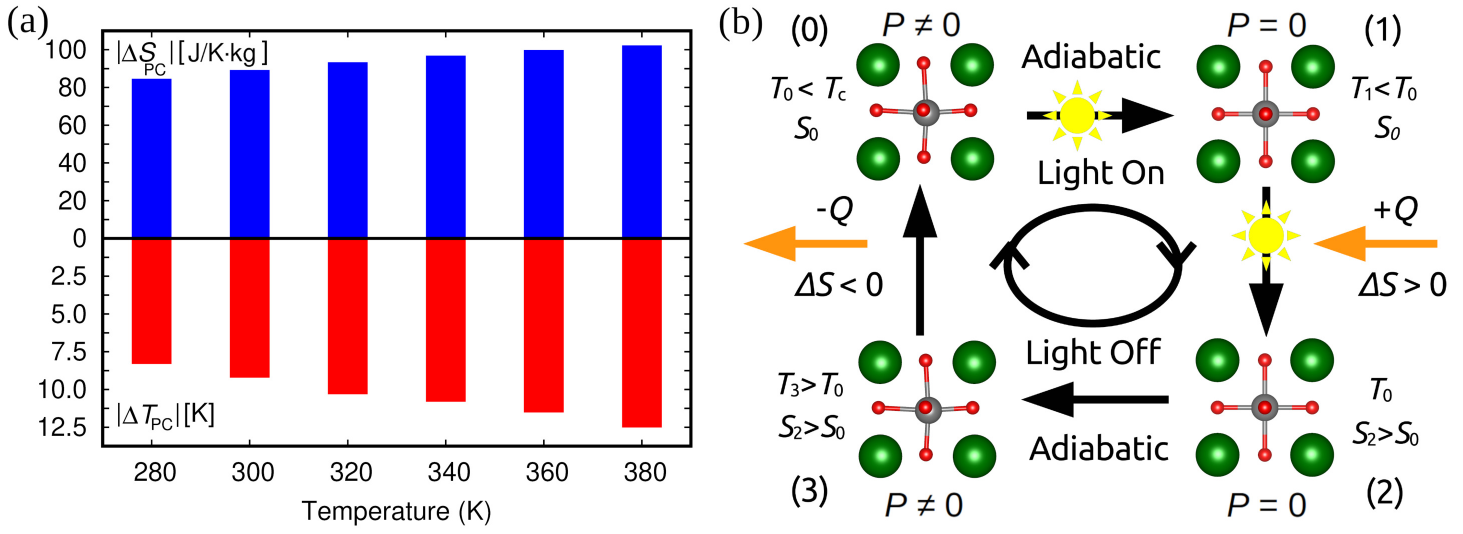}}
\caption{\textbf{Extremely ample photocaloric operation conditions and envisaged solid-state refrigeration cycle.} 
	(a)~The magnitude of the two \textit{photocaloric} descriptors $|\Delta S_{\rm PC}|$ and $|\Delta T_{\rm PC}|$ 
	evaluated for KNbO$_{3}$ as a function of temperature. (b)~A possible four-step solid-state refrigeration cycle 
	based on the \textit{photocaloric} effect unveiled in this study.}
\label{fig5}
\end{figure*}

Figure~\ref{fig4} encloses examples of our evaluations for $\Delta S_{\rm PC}$ and $\Delta T_{\rm PC}$, considering different 
temperature intervals in which the polar $R$ and $O$ phases are observed to be thermodynamically stable. For instance, at a 
temperature of $\approx 220$~K and considering the $R$ phase under dark conditions, we estimated a giant isothermal entropy 
change of $89.5$~J~K$^{-1}$~kg$^{-1}$ and an adiabatic temperature change of $-8.0$~K ($\Delta T_{\rm PC} < 0$ since the
state stabilized with light has higher entropy). Similarly, at room temperature and considering the $O$ phase under dark 
conditions, we computed a substantial isothermal entropy change of $89.2$~J~K$^{-1}$~kg$^{-1}$ and an adiabatic temperature 
change of $-9.2$~K. These \textit{photocaloric} results are highly promising for technological applications as they are 
comparable in magnitude to those obtained from state-of-the-art electrocaloric (EC) effects \cite{ec1,ec2,ec3,ec4,ec5}. 
Furthermore, the magnitude of the $|\Delta T_{\rm PC}|$ predicted for KNO is approximately one order of magnitude larger 
than the EC adiabatic temperature changes measured for similar polar materials such as BaTiO$_{3}$ (BTO) \cite{btoec}. The 
relatively large $|\Delta T_{\rm PC}|$ as compared to $|\Delta T_{\rm EC}|$ (or equivalently, $|\Delta S_{\rm PC}| > 
|\Delta S_{\rm EC}|$) can be attributed to the unique nature of the light-induced phase transition: the nonpolar $C$ state
is directly stabilized, bypassing the intermediate polar states that are typically encountered during conventional $T$-induced 
transformations \cite{wang20,shirane54}. Consequently, polar materials similar to KNO and presenting high Curie temperatures 
are probably most suitable for achieving substantial \textit{photocaloric} responses.

A highly promising aspect of the \textit{photocaloric} effect theoretically unveiled in this study, which sets it apart from 
other known caloric effects, is its operability over vast temperature intervals defined by the ranges of thermodynamic stability 
of the polar phases of the material. This exceptional characteristic is explicitly demonstrated in Fig.~\ref{fig5}a, where we 
represent the values of the $|\Delta S_{\rm PC}|$ and $|\Delta T_{\rm PC}|$ estimated for KNO over an unprecendented wide 
temperature range spanning more than a hundred Kelvin, encompassing room temperature. Interestingly, both the isothermal entropy 
and adiabatic temperature changes slightly but steadily increase with increasing temperature. For example, at $T = 380$~K, 
these quantities amount to $102.2$~J~K$^{-1}$~kg$^{-1}$ and $12.50$~K, respectively. This behavior arises from the fact that 
the variation of the entropy upon increasing temperature, or equivalently the heat capacity $C_{v} = \frac{1}{T} \left( 
\partial S / \partial T \right)_{V}$, is somewhat higher for the nonpolar $C$ phase than for the polar $O$ phase (Fig.~\ref{fig4}). 
The likely persistence of PC effects at very low temperatures makes the proposed scheme an appealing alternative for 
achieving cryogenic temperatures, which can be particularly useful in applications such as quantum computing. This possibility 
warrants future detailed theoretical exploration due to the inevitable consideration of quantum nuclear effects \cite{qha}.

A possible four-step refrigeration cycle based on the \textit{photocaloric} effect observed in KNO is envisioned 
(Fig.~\ref{fig5}b). 
$(0) \to (1)$~\textit{Adiabatic light exposure:} Initially, starting from the polar low-entropy $O$ phase at room 
temperature, $T_{0}$, the KNO system is adiabatically irradiated until stabilizing the nonpolar high-entropy $C$ phase. 
As a result, the temperature of the KNO sample decreases, $T_{1} < T_{0}$ ($\Delta T_{\rm PC} < 0$).
$(1) \to (2)$~\textit{Heat transfer:} With the light conditions maintained, the KNO sample is placed in contact with the 
targeted body to be refrigerated, facilitating the transfer of heat towards KNO, $| Q | = T \cdot |\Delta S_{\rm PC}|$, 
until restoring its initial temperature conditions, $T_{2} = T_{0}$.
$(2) \to (3)$~\textit{Adiabatic light removal:} The light is adiabatically switched off, stabilizing the polar 
low-entropy $O$ phase again. Consequently, the temperature of the KNO sample increases, $T_{3} > T_{0}$.
$(3) \to (0)$~\textit{Heat removal:} Finally, the KNO sample is placed in contact with a heat sink, allowing heat to flow 
from KNO until restoring the initial temperature conditions, thereby completing a cooling cycle.

In conclusion, we have presented compelling theoretical evidence for the existence of giant \textit{photocaloric} effects 
driven by light absorption in the archetypal ferroelectric perovskite KNbO$_{3}$. The magnitude of the unveiled \textit{photocaloric} 
effects is remarkably large, that is, comparable to the most promising electrocaloric effects measured in ferroelectrics. 
An unparalleled and distinctive feature of these \textit{photocaloric} effects is that they remain very large over vast 
temperature intervals spanning several hundred Kelvin, only limited by the ranges of thermodynamic stability of the polar 
phases of the material. Therefore, this study provides motivation for exploring new concepts and strategies to develop 
environmentally friendly and highly miniaturizable solid-state cooling technologies.

\section*{Acknowledgements}
We acknowledge financial support by the Spanish Ministry of Science under the grants PID2020-119777GB-I00, TED2021-130265B-C22, 
PID2020-112975GB-I00, RYC2018-024947-I and CEX2019-000917-S, and by the Generalitat de Catalunya under the grants 2021SGR-00343 
and 2021SGR-01519. Computational support was provided by the Red Española de Supercomputación under the grants FI2022-1-0006, 
FI-2022-2-0003 and FI-2022-3-0014. We thank J. Carrete for useful discussions.

\end{document}